\documentclass{aa}  
\newcommand{\kms}{km s$^{-1}$} 

\usepackage{graphicx}
\usepackage{txfonts}
\begin{document}

   \title{The incidence of binaries in Globular Cluster stellar populations\thanks{
Based on data obtained with the Very Large Telescope at the European Southern 
Observatory, programs: 073.D-0100, 073.D-0211 and  083.D-0208}}


   \author{S. Lucatello\inst{1,2}, A. Sollima\inst{3}, R. Gratton\inst{1}; 
E. Vesperini\inst{4}, V. D'Orazi\inst{1,5,6}, 
E. Carretta\inst{3} and A. Bragaglia\inst{3}
     }

   \institute{INAF -- Osservatorio Astronomico di Padova, Vicolo dell'Osservatorio 5, 
35122 Padova, Italy.\\
              \email{sara.lucatello@oapd.inaf.it, raffaele.gratton@oapd.inaf.it, valentina.dorazi@oapd.inaf.it}
		\and
		Visiting scientist, European Southern Observatory, Garching, Germany 
		\and
	       INAF --  Osservatorio Astronomico di Bologna, via Ranzani 1, 40127 Bologna, Italy\\
	\email{antonio.sollima@oabo.inaf.it, eugenio.carretta@oabo.inaf.it, angela.bragaglia@oabo.inaf.it}
	\and
	   Department of Astronomy, Indiana University, Bloomington, IN 47401, USA\\
             \email{evesperi@indiana.edu}
	  \and
	Monash Centre for Astrophysics, School of Physics and Astronomy, Monash University, Clayton, VIC 3800, Australia 
	\and 
	Department of Physics and Astronomy, Macquarie University, North Ryde, NSW 2109, Australia
             }

   \date{}

 
  \abstract{
Binary fraction and orbital characteristics provide indications on the conditions of star formation, as they shed light on
 the environment they were born in. Multiple systems are more common in low density environments rather than in higher density ones. 
In the current debate about the formation of Globular Clusters and their multiple populations, studying the 
binary incidence in the populations they host offers a crucial piece of information on the environment of 
their birth and their subsequent dynamical evolution.

Through a multi-year observational campaign using FLAMES at VLT, we monitored the radial velocity of 
968 Red-Giant Branch stars located around the half-light radii in a sample of 10 Galactic Globular Clusters. 
 We found a total of 21 radial velocity variables identified as {\it bona fide} binary stars, for a binary fraction 
of 2.2\%$\pm$0.5\%. When separating the sample into first generation and second generation stars, 
we find a binary fraction of 4.9\%$\pm$1.3\% and 1.2\%$\pm$0.4\% respectively. Through simulations
that take into account possible sources of bias in detecting radial velocity variations in the two populations, 
 we show that the difference is significant and only marginally affected by such effects. 

Such a different binary fraction strongly suggests different conditions in the environment of formation 
and evolution of first and second generations stars, with the latter being born in a much 
denser environment. 

Our result hence strongly supports the idea that the second generation forms in a dense subsystem 
at the center of the loosely distributed first generation, where (loose) binaries are 
efficiently destroyed. }
 

\authorrunning{Lucatello et al.}

   \maketitle
%

\section{Introduction}
Strong evidence has accumulated in the recent years supporting the
concept that globular clusters (GCs) are composed of different stellar
populations, characterized by differences in their chemical
compositions (for reviews, see Gratton et al. 2004 and Gratton et
al. 2012). In a typical globular cluster, about a third of the stars
have element-to-element abundance ratios that are indistinguishable
from those typically observed in field metal-poor stars having similar
[Fe/H] values. However, the remaining (majority) stars show
enhancements of some elements (N, Na, Al) and depletion of others (O,
Mg) that can be attributed to high temperature H-burning, while
e.g. Fe-peak element abundances are very nearly the same in all stars.
This abundance pattern is seen among all evolutionary phases \citep[see][and references therein]{gratton12}.

This led most authors to think of a multiple generations scenario where
the stars with peculiar composition (second generation, SG) formed from
material lost at low velocity by a fraction of the stars with normal
composition \citep[first generation, FG; see e.g.,][]{ventura01,decressin07}.
Recently, \citet{bastian13} proposed a
scenario where the stars with peculiar composition actually formed
together with those with normal composition, but accreted (quite a
large) fraction of their mass from the ejecta of massive
binaries. 
While in this last scenario
there are not two distinct stellar generations, it predicts the presence
of stellar populations differing in chemical composition. For simplicity,
throughout this paper we will still use the terminology "first/second
stellar generations" to distinguish between stars with primordial/altered
chemical composition, respectively, although the real presence of
different stellar generations is still under debate and one 
could refer to these two groups as pristine and enriched stars.
 \footnote{Even more recently however 
Bastian et al. (2015) have argued that no self-enrichment scenario  can explain the full variety 
of phenomena observed in GCs}.

First and second generation stars in globular clusters do not differ uniquely for their
chemical composition. Several studies have demonstrated that at least
in some clusters, the second generation stars are more centrally
concentrated than the first generation ones 
(see e.g., \citet{sollima07,lardo11,milone12,kucinskas14};
but see also \citet{larsen15} 
for a study of M15 showing that the primordial population fraction decreases from the outskirts to the half-mass radius and then increases again towards the very center). 
The
correlation between chemical and dynamical properties may shed light
onto the same mechanism of formation and evolution of the
clusters \citep[]{dercole08,henault15}, in spite of the fact that they are
very old objects. 

The largest concentration of second generation stars suggest that they
formed in higher density regions. On the other hand, two-body relaxation
occurring in the long-term evolution of collisional systems like GCs tend
to erase structural differences imprinted in the early stages of
formation of these objects \citep{decressin08,vesperini13}.

Binary stars are the ideal tool to reveal the signature of
primordial differences in the environment where FG and SG formed. 
Only a fraction of the primordial binaries are expected to survive to the
dynamical processes occurring during the evolution of a GC. The main
process affecting the binary fraction is the ionization through collisions
with single stars and other binaries whose efficiency depends on the
environment density and velocity dispersion. So, the trace left by
differences in the primordial environment where FG/SG binaries formed on
their fraction and period distribution remain frozen and can still be
visible today (Vesperini et al. 2011; Hong et al. 2015). 
In fact, while very wide binaries are destroyed
even in environments with moderate density, and the closest binaries
are expected to survive even at high densities, the destruction rate
of those of intermediate separation - with binding energy comparable
to the typical kinetic energy of cluster stars - is expected to
strongly depend on the encounter rate and then on the density of their
environment. We might then expect systematic differences in the
fraction of this type of binaries among different generations. While
there are several complicating factors, like for example the segregation of
binaries toward center of the clusters due to energy equipartition
or the formation of new binaries in three-body encounters, it should 
still be possible  to detect systematic differences in the frequency of binaries
in stars of different stellar generations by comparing samples of
stars of the different stellar generations observed at similar
distances from the cluster center.  

Studies of the fraction of
binaries in clusters have been presented by several authors ,
based on both variation of radial velocities (RVs hereafter, see e.g.,
 \citealt[]{ym94,yc96,yr96,albrow01, sommariva09}), mainly in
giants, and from photometry, mainly for main sequence stars 
(see e.g., \citealt[]{sollima07,rubenstein97,ji13,dalessandro11}). 
Milone et al. (2012) presented a very extensive photometric study
based on the HST ACS survey, showing that the binary
fraction is generally quite low in globular clusters and it is a
function of the cluster absolute magnitude and then likely of the
cluster mass: massive clusters have very few binaries, while the
binary fraction in smaller clusters is much closer to the value
observed in field stars and open clusters. 
While this result is of high interest, 
we are however not yet able to tag the stars observed by Milone et al. as
members of different stellar generations, though this will perhaps be
possible in the future through the analysis of the extensive photometric data gathered with
the HST  (see e.g. Piotto et al. 2015). 

To date, the largest homogeneous sample of GC stars with high resolution spectroscopy is that 
collected within the FLAMES Globular Cluster Na-O survey by Carretta et al. (2015 and references 
therein) . 
Fe, O and Na abundances have been measured for these objects, 
allowing to classify them into first and second
generation stars. The spectra also provide first epoch RVs. In the last few years, we gathered additional second epoch
RVs for a significant fraction of these stars, allowing
to measure RV variations possibly indicating their
membership in binary systems. 

D'Orazi et al. (2010) studied the incidence of Ba stars in GC populations, 
finding that they are much more common among the so called 
primordial population (essentially equivalent to the FG) than in the 
Intermediate and Extreme components (which make up the SG). 
This finding provided evidence of a different binary fraction in the GC populations.
In fact, the abundance pattern characteristic of Ba stars originate from mass transfer in a
binary from a companion of $\sim$1.5\,M$_{\odot}$ during its Asymptotic Giant Branch 
phase. So while their incidence does not measure the 
binary fraction of the parent population
itself, it does trace it indirectly. 
In the same paper D'Orazi et al. also presented an RV based study  
of the binaries in NGC6121, obtained a relatively large frequency of
binaries for first generation stars, and only a low upper limit for
the second generation ones, indicating a large difference between the
two populations.

\section{Observations and data analysis \label{obs_sec}}
Data considered are a combination of archival and proprietary
ones collected with the ESO Very Large Telescope at Cerro Paranal 
using GIRAFFE-FLAMES (Pasquini et al. 2004).  Proprietary observations were obtained within ESO programs
072.D-0507, 073.D-0211, 083.D-0208, 085.D-205 and 088.D-403.  Programs
072.D-0507, 073.D-0211 and 083.D-0208 targeted 19 GCs with the aim of
characterizing the Na-O anti-correlation in a large sample of GCs
\citep[see][]{carretta09}.  Observations were collected using the HR11 and
HR13 setups, which yield spectra in the range 5600---5840\,\AA~with
R$\simeq$24,200 (appropriate for the measurement of Na abundance) and
6120---6405\,\AA~with R$\simeq$22,500 (for O abundances) 
respectively.  Due to the choice of maximizing the
number of targets observed with the UVES fibers, different exposures
have slightly different fiber positionings, however most stars are
observed (typically several times, see Table
\ref{tab_obs} for details) with each of the setups. 
For further details about the observations and the adopted 
target selection criteria the interested reader is referred to 
Carretta et al. (2009).

Program 085.D-205 collected spectra with the HR21 setup (8757--9001\,\AA~and
R$\sim$16,200) to derive Al abundances (using the same positionings)
in 4 of the clusters (NGC104, NGC1904, NGC6121, NGC6752) involved in
the Na-O survey.  For one of the clusters, NGC7078, we collected
archival data (ESO program 080.B-0784) taken using the same
fiber positionings as in our programs, using the HR14 setup,
(6300--6690\,\AA, R$\simeq$17,700).  Finally, we collected data for 6
of the clusters (NGC104, NGC288, NGC1851, NGC3201, NGC5904, NGC7099) 
using the HR9 setup (5143--5356\,\AA, R$\sim$25,900;
program 088.B-403), centered on the \ion{Mg}{I} b triplet.  With the exception
of the last one, the primary scientific aim of the different projects
were abundance analysis, hence the temporal baseline and spacing of
the observations is far from ideal for a systematic search of RV variables.

The combination of this dataset provides multi-epoch observations with
a baseline of at least 3\,yr for 10 GCs, namely: NGC104 (47Tuc), NGC1851, NGC1904, 
NGC288, NGC3201, NGC5904, NGC6121\footnote{Data used in D'Orazi et al. (2010) 
are also included in our dataset} , NGC6752,  NGC7078 and
NGC7099.  Atmospheric parameters, Fe,
Na and O abundances are measured for all the stars in the present
sample \citep[see][and references therein]{carretta09,carretta11}. Table
\ref{tab_obs} shows the details of the observations.
\begin{table}
\begin{center}
\caption{Observational log \label{tab_obs}.Table is provided in its entirety in electronic version.}
\begin{tabular}{lrrr}
\hline\hline
Cluster ID & MJD& setup  & Exposure \\
&&&Time (s)\\
\hline
NGC104 &53182.41 & HR11 & 1600 \\
NGC104 &53193.38 & HR11 & 1600 \\
NGC104 &53193.41 & HR13 & 1600 \\
NGC104 &55407.33 & HR21 & 2790 \\
NGC104 &55845.12 & HR9 & 1115 \\
\hline
\end{tabular}
\end{center}
\end{table}

Data were reduced with the standard ESO FLAMES GIRAFFE pipeline
(different versions were used: 2.5.1 and 2.5.3, however for the purposes
of the present analysis it does not make any difference, given that possible 
systematic shifts in RV introduced by differences in 
adopted wavelength calibrations are accounted for, see below in this section).  Continuum
fitting, sky and telluric spectra subtractions were performed in
IRAF\footnote{IRAF is distributed by the National Optical Astronomy
Observatories, which are operated by the Association of Universities
for Research in Astronomy, Inc., under cooperative agreement with the
National Science Foundation.}.

For each cluster and each setup we selected the highest S/N
spectrum, whose RV was estimated by identifying several dozens of
spectral lines and measuring their shifts. These spectra were then
brought to rest-frame and then used as templates in
the subsequent cross-correlation (performed with the IRAF task {\it
fxcor}) with which RVs for the rest of the stars in the clusters were
measured. Appropriate barycentric corrections for the different
spectra were applied to derive heliocentric RVs.  Those
derived from spectra taken within a 24\,hr period were averaged and
considered as a single RV point.  Before doing so, we checked for RV
shifts between the exposures taken within such period and found them to be negligible for
all cluster members stars in all cases.

We note that in the case of very metal poor clusters, such as NGC7078
and NGC1904, even combining exposures taken less than one day apart,
errors on a single RV determination (which also takes into account the
error introduced when shifting to a common RV system, see below in
this section) for a star are quite large, ranging from $\sim$0.5 to
over 2 \kms, while for more metal rich clusters, whose stars have
spectra characterized by a large number of features, the error is
typically much smaller, $\sim$0.3\,\kms.

In order to check for systematic offsets between RVs measured from
spectra collected with different setups and/or at different times and reduced 
with different version of the pipeline, we determined a 3\,$\sigma$-clipped mean (after discarding outliers
flagged from chemical composition as in Carretta et al. 2009) for the RV in each exposure. 
Those mean values from each exposure in each setup and for each cluster were then compared, 
The derived offsets are quite small, $\sim$0.2\,\kms~between spectra taken with the same
setups at different times, while they are typically larger between
different setups, as much as $\sim$3\,\kms. This suggests the
presence of systematic offsets, which might hamper the derivation of
the correct RV and the detection of RV variables.
This offset between different setups has been noted by others, see e.g. the cases discussed by
the Gaia-ESO Survey (Lardo et al. 2015).

To address this problem, we proceeded as follows.  For each cluster we
chose as reference frames the RV determined from spectra obtained with
the HR11 setup. If multiple HR11 exposures existed the one with the
highest S/N ratio was selected.  The reason behind the choice of this
setup is that HR11, with HR13, is the only setup available for all
clusters and the RV measured from the HR11 spectra have typically
smaller errors than those from the HR13, especially for metal poor
clusters. This could be due to the fact that the spectral region
covered with this setup has more detectable lines and less
contamination from telluric lines, whose subtraction residuals can
lead to larger uncertainties in the RV determinations via
cross-correlation.  
For each frame a correction was calculated
(accounting for the differences in the mean RV in that frame with
respect to that of the reference frame) and applied to the individual
RVs. This procedure brings all the RVs to a common system and
minimizes the effects of instrumental/data reduction offsets in RV
variable searches.  Note that the adopted RV errors on the individual
stars account for the errors due to the application of such shifts,
which are typically larger in clusters with fewer stars observed.

Due to the already discussed change of fiber positioning between
different exposures, not all stars were observed in every single
exposure. Moreover because of the quite wide variety of the extension
and concentration of the clusters in our sample, due to FLAMES fiber
collision limitations, the number of stars observed at more than one
epoch for each clusters varies from 44 (NGC7099) to 150 (NGC6752) (see
Table \ref{stars_tab} for details). Table \ref{stars_tab} lists also 
the derived average RVs, 
which are in excellent agreement with those published 
in the Harris (1996) catalog, while the derived velocity dispersions are generally 
smaller, which is what is expected given that Harris lists the central velocity dispersion, 
while our observations target stars around half-light radius. 
\begin{table*}
\begin{center}
\caption{Number of stars considered. 
Columns two to five lists stars with more than one RV epoch, while the the fifth 
all the stars with at least one RV measurement. Labeling 
of FG and SG according to Carretta et al. (2009).  \label{stars_tab}}
\begin{tabular}{lrrrccccc}
\hline\hline
Cluster & Object   & FG  & SG & Unlabeled & Object &$<$RV$>$ & $\sigma_{<RV>} $& $\sigma_{RV}$\\
      ID         &     \#            &        & &      &  \#& km s$^{-1}$  &  km s$^{-1}$  &  km s$^{-1}$   \\
\hline
NGC104   &121   & 39    & 82   &  0 & 148  &  -18.0  &  9.6  &  0.8\\
NGC1851 & 116  & 35     & 80  &  1  & 117  &  321.5  &  3.5  &  0.3\\
NGC1904 &  53   &  17   & 26   & 10 &  53  &  206.7  &  3.3  &  0.5\\
NGC288   &  108 &  43   &  64  &  1  &108  &  -44.6  &  2.8 &  0.3\\
NGC3201 & 101  &  43   & 57   &  1  & 149  &  498.8  &  3.4  &  0.3\\
NGC5904 & 113   &  31   &  82  &  0  & 136  &  51.5  &  5.1 &  0.4\\
NGC6121 &   85  &  29    &  56  & 0   & 103  &  71.4  &  3.8  &  0.4\\
NGC6752 & 150  &  27    &  99  &  24 &151  &  -27.3  &  6.4  &  0.5\\
NGC7078 & 77    &  13    &  32  & 32  & 77  &  -107.3  &  5.3  &  0.6\\
NGC7099 & 44    &  11    &  26   & 7   & 64  &  -185.0  &  2.9  &  0.4\\
\hline
Total & 968 & 288 & 604 &  76     &       1106\\
\hline
\end{tabular}
\end{center}
\end{table*}

\citet[]{preston01} found that velocity errors derived from multiple 
observations of constant RV metal-poor giant stars are larger 
than the error in RV from individual spectra by a factor of
$\sim$2–-3.  It is known that some metal-poor red giants exhibit
velocity jitter to the level of $\sim$1.5-2\,\kms \citep[]{carney03}.
This phenomenon seems to affect only the intrinsically brightest stars,
within $\sim$0.5--1 mag from the red giant branch tip and is unfortunately
indistinguishable from RV variations due to binarity.  In
five out of the ten clusters in our sample there are stars that meet
this criterion (1 mag from the red giant tip), for a total of 17 objects.  To ensure that our results
are not influenced by the RV jitter, we have excluded those stars from
our sample.

To determine the probability of RV variations, we follow
the procedure described in \citet[]{lucatello05}. We calculated for each 
star the $\chi^2$ value for the RV distribution and then 
evaluated the the probability Q($\chi^2$|$f$) that the RV values derived 
for each star are not compatible with different measurements of the same 
quantity, i.e. the probability that the observed scatter is due an intrinsic 
change in the measured quantity rather than to experimental errors. 
%

We consider as a positive RV variable identification a
value of Q $>$ 0.995 (i.e. a probability that the observed RV
variations arise from observational scatter P=1-Q$<$0.005). Our
criterion is very restrictive and it is likely to fail to detect a non
negligible fraction of binaries in our sample. However, since our aim
is to assess binarity ratio in a strictly differential way between the
two populations (see next section) we prefer to minimize the number of
false detections (which scale with P) rather than reaching higher
completeness.
Results are reported in Table \ref{prob_tab}.

\begin{table*}
\begin{center}
\caption{Probability result for each star in the sample. Stellar generation label (FG/SG), average RV, $\sigma$, $\chi^{2}$, degrees of freedom, probability of RV variation, and [Na/Fe]. Table is given in its entirety in on line version. \label{prob_tab}}
\begin{tabular}{lcccccccr}
\hline\hline
Cluster ID & Star ID & Fg/SG & $<$RV$>$   & $\sigma_{<{\rm RV}>}$ &$\chi^2$ & f & Q & [Na/Fe]\\
&&&(km s$^{-1})$&(km s$^{-1})$&&&&(dex)\\
\hline
    NGC104   &    2608   &  SG   &    -27.63   &      0.63   &     2.574   &     3   &     0.538   &     0.615   \\
    NGC104   &    2871   &  SG   &    -21.78   &      0.30   &     0.283   &     2   &     0.132   &     0.440   \\
    NGC104   &    4373   &  FG   &    -12.40   &      0.47   &     1.523   &     3   &     0.323   &     0.249   \\
    NGC104   &    5172   &  SG   &    -19.37   &      0.43   &     1.362   &     3   &     0.286   &     0.489   \\
    NGC104   &    5277   &  SG   &    -19.48   &      0.36   &     0.862   &     3   &     0.165   &     0.423   \\
    NGC104   &    5640   &  FG   &     -9.96   &      0.43   &     1.245   &     3   &     0.258   &     0.215   \\
    NGC104   &    6092   &  FG   &     -5.41   &      0.07   &     0.013   &     2   &     0.007   &     0.307   \\
    NGC104   &    6808   &  SG   &    -12.35   &      0.48   &     1.562   &     3   &     0.332   &     0.421   \\
    NGC104   &    7711   &  SG   &    -13.55   &      0.43   &     1.296   &     3   &     0.270   &     0.426   \\
    NGC104   &    7904   &  SG   &    -28.14   &      0.33   &     0.726   &     3   &     0.133   &     0.438   \\
    NGC104   &    9163   &  SG   &    -11.94   &      0.19   &     0.120   &     2   &     0.058   &     0.526   \\
    NGC104   &    9518   &  FG   &    -16.59   &      0.44   &     1.369   &     3   &     0.287   &     0.256   \\
\hline
\end{tabular}
\end{center}
\end{table*}

\section{Discussion \label{disc_ref}}

\begin{figure*}
\includegraphics[angle=0, width=18cm]{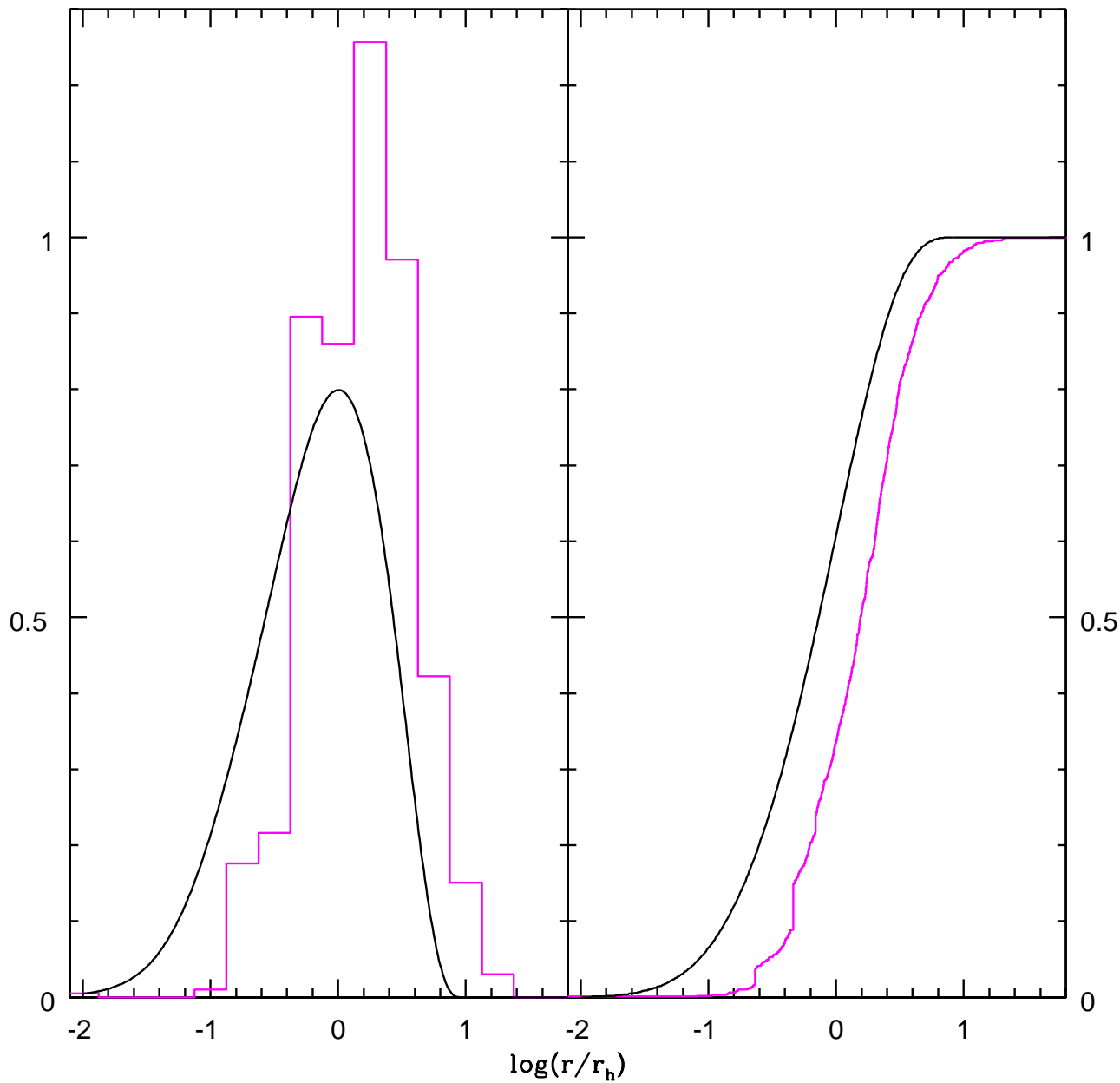}
\caption{Left panels: distribution of mass as a function of radius (expressed in half-light radius units) for our sample clusters (black) and distribution of stars observed in this study (magenta). Right panels: cumulative distribution for the mass of our sample GCs (black) and 
of the stars observed in this study (magenta).}
\end{figure*}
\begin{figure*}
\includegraphics[angle=0, width=18cm, height=20cm]{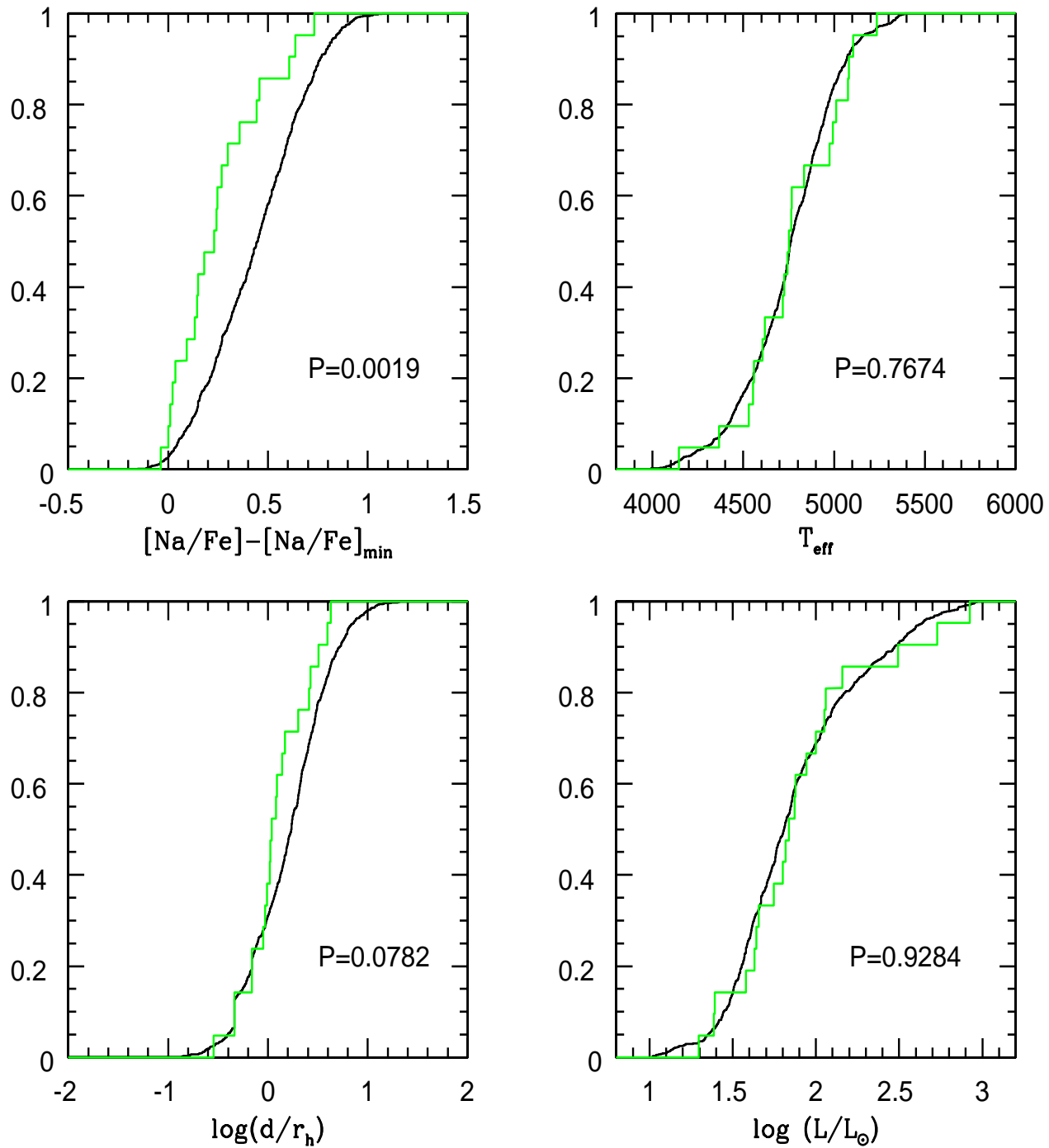}
\caption{Cumulative distribution of distance from the center (scaled with respect to half-light radii), luminosity, 
effective temperature and Na enhancement (see text for definition) for {\it bona fide} binary (green) and single (black) stars, 21 and 931 objects respectively. The p-value for two sided Kolmogorov-Smirnov test is also reported. \label{fig_binsin}}
\end{figure*}
\begin{figure*}
\includegraphics[angle=0, width=18cm, height=20cm]{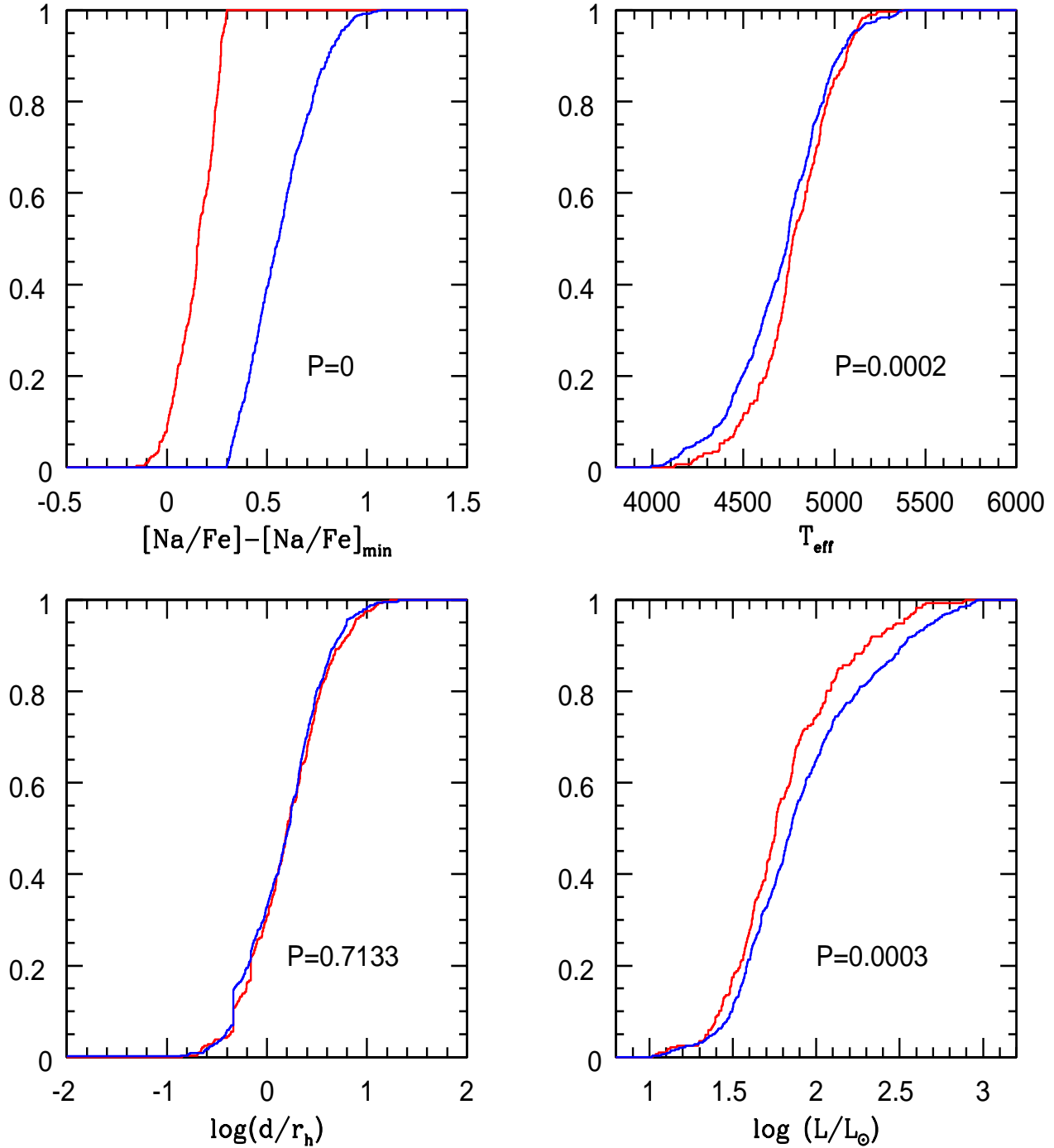}
\caption{Same as Fig. \ref{fig_binsin} for FG (red) and SG (blue) objects.  \label{fig_fgsf}}
\end{figure*}

 We note that while the clusters in our sample have in fact quite
different characteristics in terms of concentration and extension,
our observations are typically concentrated around the half-light radius i.e. the distance
typical of the bulk of the stellar mass in a cluster. In Figure 1, the
distribution of distances from the cluster center (in units of half-mass
radii; from the Harris catalog) of our observed sample is compared with
that predicted for the entire cluster population. To compute this last
quantity the mass profiles of clusters in our sample have been calculated
from their best-fit King (1966) model, normalized to the number of targets
and summed together. The two distributions are 
similar, with ours being slightly shifted towards more central stars, 
indicating that our sample is a fair representation of the entire
cluster population. Therefore, conclusions derived for the stars in our
sample do apply to the bulk of the stars in the clusters.

The sample of GCs observed in the present study is quite heterogeneous in terms of metallicity, mass and
horizontal-branch morphology and it would be of high interest to study 
the binary fraction as a function of these characteristics. However, 
small number statistics hampers our capability of deriving a reliable binary 
fraction in each of the clusters and investigating any correlations of these
characteristics with binarity. Such exercise will thus not be attempted and the discussion
will be limited to the combined sample, which yields robust results.

Out of a grand total of 968 objects, 21 show evidence of RV variation
at the 0.995 confidence level, for an overall binary fraction of
2.2\%$\pm$0.5\%. We note that this fraction cannot be simply compared
to the values found with other methods and/or other evolutionary stages 
(see e.g. Milone et al. 2012, Ji \& Bregman 2013 and 2015, Moni Bidin et al. 2011) as it
is established purely from RV variations measured in RGB stars with our particular observing 
pattern and RV precision, which is typically effective 
in detecting binaries with periods shorter than $\sim$10 years (see section 4).

Moreover, it is limited to a very partial area within each
clusters (due to the FLAMES field size) typically around the
half-light radius, where the binary fraction is likely different from the cluster 
center.

 Figure 2 shows the cumulative distributions of a few characteristics of
 the {\it bona fide} binaries and single stars, which also shows the 
p-value for a two-sided Kolmogorov-Smirnov test for each of the distributions.
We see no significant difference in effective temperatures and 
luminosity distributions. 
There is no strong indication of a difference in distribution as a function of
distance from  cluster center (scaled with respect to the half light radii of the respective cluster). 
Lack of such finding does not exclude the presence of such an effect, 
given the limited range in radius spanned by our study.
On the other hand, the distributions of 
the Na enhancement (defined for each star as the difference between its 
[Na/Fe] and [Na/Fe]$_{min}$ for its host cluster as listed in Carretta et al 2009 and 2010), 
are quite different for {\it bona-fide} single and binary stars, with the 
latter having a typically lower Na content, in line with that typically 
found among FG stars.

In fact, when considering the binary fraction in the two populations, we note that 
they are quite different. Our sample includes 288 FG and 604 SG stars, 
with 14 and 7 binaries, respectively. 
Note that the number of stars in the combined sample is larger than the sum of FG and SG 
because for 76 objects Na abundances, which is required to label a star as belonging to 
one of the populations, could not be measured as the appropriate spectral range was not observed.
The binary incidence in the two population is hence 
4.9\%$\pm$1.3\% among FG stars  and 1.2\%$\pm$0.4\% among SG stars.

Such difference is quite striking. While we stress again that these values 
are by no means an accurate estimate of the overall binary fraction among these 
population, the ratios of the two fractions is a much more robust quantity.
In fact, observations, reduction and analysis were performed identically, regardless of 
which population the object belonged to. 

It is interesting to compare this results to the findings of D'Orazi et al. (2010). 
They reported an overall Ba fraction of $\sim$0.4\%, and an incidence among FG stars of 
$\sim$2\%, to be compared to our finding an overall binary fraction 
of 2.2\%$\pm$0.5\% and  4.9\%$\pm$1.3\%
respectively. While we should keep in mind that small number statistics 
do play a role (out of 1205 stars they find 5 Ba stars, 4 of which are FG), 
these numbers are consistent with Ba stars being a special case 
among binaries.  

When  considering the binary fraction in NGC6121, they reported an overall fraction of 
$\sim$5\%, $\sim$1\% for SG and $\sim$12\% for FG, which is a larger difference than we find in the 
overall sample. It should however be kept in mind that the present analysis 
adopts a procedure that minimizes the systematic effects on RV measurements 
due to using different set-ups and adopts a much more restrictive criterion to identify a star 
as a binary, making the results more robust.

There are however a number of biases that might in principle affect our findings.
Figure 3 shows the same distributions as Fig 2, but for FG and SG objects.
Once again, there is no evidence of difference in radial distribution, which 
has however scarce significance as discussed in the case of single vs binary stars.
Beside the obvious difference in the Na enhancement distributions, 
both effective temperature and luminosity distributions show marked differences 
in the two populations. This is not surprising: SG stars are expected to have 
an enhanced He content with respect to FG stars, and hence a slightly brighter magnitude.
(see Bragaglia et al. 2010) and higher luminosity ($<\Delta \log$ (L/L$_{\odot})>=0.11\pm0.03$). 
Given that the method followed to determine 
T$_{eff}$ in the present sample is based on the derivation of a temperature-magnitude relation, 
the difference in the temperature ($<\Delta $T$>=63\pm15$\,K) distribution follows.

Such difference in luminosity should result in spectra of typically higher signal-to-noise 
ratios for SG stars than for FG stars, which translate into smaller errors on the 
derived RVs. However, because of the non-uniformly accurate fiber positioning, 
the signal-to-noise ratio is not related just to magnitude, but depends also on 
position on the field.
Other factors that can play a (smaller) role are the abundance enhancement and 
depletion. In fact, spectra of SG stars have stronger Na lines than FG stars of similar 
atmospheric parameters, which produce stronger signatures in the cross-correlation and
lower errors in RVs derived from HR11 spectra. 
On the other hand, SG stars have often depleted Mg abundances, yielding higher errors in the RV measured 
from HR9, which contains the Mg b triplet. A simple direct assessment of these biases is hard to quantify, however simulations 
can provide an estimate of the completeness of the binary detections in the 
two populations.


\section{Simulations and completeness}
To estimate the completness of both samples of FG and SG stars
we simulate a synthetic population of binaries and applied the same technique
described in Sect \ref{obs_sec} to detect their velocity changes. 
For this purpose, for each observed target, a sample of $10^{3}$ synthetic 
binaries have been simulated by assuming
a mass of the primary component of $m_{1}=0.8 M_{\odot}$ (typical of a GC RGB star)
and randomly extracting a secondary component ($m_{2}$) from a flat distribution (Milone et al. 2012) 
between 0.1 and 0.8 $M_{\odot}$. Periods ($P$) and orbital eccentricities ($e$) 
have been assigned following the prescriptions of Duquennoy \& Mayor
(1991) and the corresponding semi-mayor axes ($a$) have been calculated using the third
Kepler law. From this library of binaries we removed all those objects whose
pericenter were smaller than a minimum separation linked to the volume averaged 
Roche lobe size (see Lee \& Nelson 1988). The projected velocity of the primary 
components have been then calculated using the relation
$$v_{1}=\frac{2 \pi a sin i}{P (1+m_{1}/m_{2}) \sqrt{1-e^{2}}}
(cos(\alpha+\theta)+e cos\alpha)$$ 
where $\alpha$ is the longitude of the periastron, $i$ is the inclination angle,
$\theta$ is the phase from periastron. The distribution of the angles at the
first epoch was chosen according to their corresponding probability distributions
(Prob(i)=Prob($\alpha$)=constant; Prob$(\theta)\propto\dot{\theta}^{-1}$).
Then, we calculated the sequence of phase angles according to the
observational pattern of the associated target and the corresponding 
velocities have been derived. A velocity shift extracted from a Gaussian
function with a standard deviation equal to the target uncertainty has been 
added to mimic the observational error. Finally, the detection procedure
described in Sect. \ref{obs_sec} has been applied to the sample of synthetic 
binaries and the completeness has been estimated as the fraction of recovered
binaries. For comparison, we applied the same procedure to the sample of field
metal-poor RGB stars of Carney et al. (2003).
\begin{figure}
\includegraphics[angle=0, width=9cm]{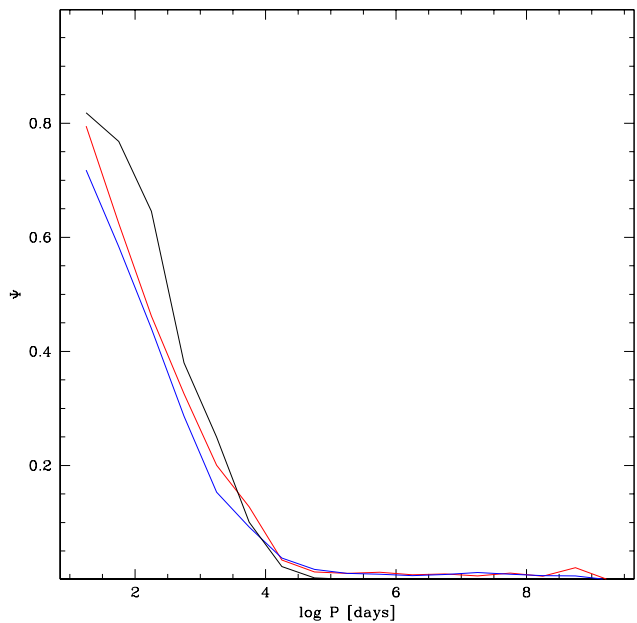}
\caption{Binary detection completeness for the Carney et al sample (in black), 
FG (blue) and SG (red) as a function of orbital period.\label{compl}}
\end{figure}

The completeness as a function of the orbital period is shown in Fig.
\ref{compl}. As expected, the completeness of all samples decrease for
increasing periods because of the smaller portion of the sampled velocity curve.
The maximum completeness at short periods ($<30$ d) reaches $\sim$70-80\%
dropping to $50\%$ at $P\sim100$ d, while at $P>10^{4}$ d no binaries are
expected to be detectable. It is apparent that the sample of FG binaries
presents a slightly larger (by $\sim3.4\%$) completeness with respect to SG. 
Such a difference, while significant, cannot explain the observed excess of 
FG/SG binaries reported in Sect. \ref{disc_ref} unless the period distribution of FG
and SG binaries were strikingly different. 
This last occurrence is however very unlikely: indeed to reproduce the detected 
ratio of FG/SG binaries one should assume that the majority of SG binaries have 
periods $>10^{4}$ d and corresponding semi-axes $>10$ AU. Such wide binaries
cannot survive in the dense environment of GCs being immediately ionized by
collisions with other cluster stars.

From the completeness derived above it is possible to check whether the fraction
of binaries in the FG is compatible with that observed in the Galactic field.
To do this, we simulated a synthetic population made by a mixture of binaries
and single stars. In this case, the period distribution of binaries has been
truncated at 6000 d (the longest time lapse in the Carney et al. sample),
corresponding to a maximum semi-axis of 7.7 AU. 

The projected velocity of binaries has been calculated as
described above, while for singles only the shift due to the velocity 
uncertainty has been added to a constant velocity. The fraction of binaries of
the synthetic population has been tuned to reproduce the same ratio in the
observed sample. In this way, both the completeness and the false detections are
accounted and the global fraction of binaries can be estimated for the different
sample of GC and field stars. In this approach we implicitly assume that FG, SG
and field binaries share the same period, eccentricity and mass-ratio
distribution. This assumption is clearly false, since only the hardest binaries
(with short periods and smaller semi-axes) are expected to survive in GCs at
odds with what happens in the Galactic field where collisions are almost absent.

On the other hand, this exercise provides a rough comparison between these
different samples. The resulting binary fractions turns out to be 15.3\%, 2.8\%
and 45.6\% for the FG, SG and field samples, respectively. The above fractions
appear in good agreement with the fractions estimated by \citet{duquennoy91}
for field stars and \citet{sollima07} and \citet{milone12} for 
GCs. It is worth noting that the above fractions depend on the adopted period upper
limit: by assuming a period distribution truncated at $P<1000$ d (corresponding
to a maximum semi-axis $a<$ 2.3 AU) they reduce to 11.6\%, 2.8\% and 18.8\% for
the FG, SG and field sample, respectively.

In any case, we conclude that the field contains a significantly larger 
fraction of binaries with respect to both FG and SG.
Note that assuming a period distribution for GCs stars shifted to shorter periods 
would result in an even smaller fraction of binaries (the detection completeness of 
our observations is bigger in case of shorter periods)
thus reinforcing this
conclusion.

\section{Conclusions \label{conc_sec}}
In this study we have presented the results of the RV 
monitoring of the largest sample to date of GC stars with measured composition. 
We have detected 21 stars which met our criteria to be identified as {\it bona fide} binary
stars, 14 of which belong to the FG (which includes 288 stars) and 7 to the SG (604 stars), 
for resulting binary fractions of 4.9\%$\pm$1.3\% and 1.2\%$\pm$0.4\% respectively.

Simulations have shown that this difference is not accountable for in terms 
of observational biases, which affect the binary detection completeness almost 
identically for FG and SG objects, nor in terms of different period distribution, 
Simulations have shown that this difference is not accountable for in terms 
of observational biases, which affect the binary detection completeness almost 
identically for FG and SG objects, nor in terms of different period distribution, 
as it would require SG to have such long periods ($10^{4}$ d) and corresponding semi-axes ($>10$ AU) 
that would imply immediate destruction at the typical cluster density.

Our findings are hence robust and provide strong evidence of an
intrinsic different binary fraction in the parent populations of FG
and SG stars. The observed difference is consistent with the results
of theoretical studies on the dynamical evolution of binaries in
multiple-population clusters \citep{vesperini10,hong15}
in which the SG forms in a dense subsystem at the center of the loose
FG early cluster as first suggested by \citet{dercole08}. 
The denser environment in which SG stars form and evolve until SG and
FG stars are completely mixed leads to an enhancement in the SG binary
disruption and evolution. While initial differences in the FG and SG
structural properties can be gradually erased during the cluster
dynamical evolution, the fingerprints of these differences can still
be visible in the different FG and SG binary fraction. \footnote{
The present  findings are not incompatible with the early disk accretion 
hypothesis put forward by \citet{bastian13}. In fact, in such scenario 
the circumstellar disks on which accretion of polluting material takes 
place have radii of $\sim$100\,AU. On the other hand, our technique is effective in 
detecting binaries that have typically smaller semi-axis, of the order of 
10\,AU, and the presence of such a closeby companion would hamper the 
formation and the stability of the circumstellar disk, preventing the accretion 
of polluting material and thus the imprinting of chemical signatures
typical of the SG on binary stars.}

While this is the most likely explanation, there are in principle
additional possibilities. The SG stars might have formed with a lower birth binary fraction because of their different composition. Variations of binary fraction as a function of
metallicity and/or composition is a very poorly studied property of
stellar population, however to date no such effect has been reported
in the literature. Differences in the FG and SG kinematical properties 
 (like e.g. anisotropy, rotation, etc.) 
in combination with differences in the spatial structure might affect the survival rate for FG and SG binaries. Finally the fact that the SG stars are, at a given time, expected to have on average slightly smaller
masses (due to the slightly shorter lifetime of He enriched stars), the SG binaries have on a average a smaller binding energy (which scales with m$^2$), resulting on a smaller binary fraction. This effect is expected to be quite small and could not explain the present results.

Finally in the context of the link between globular clusters and halo
field stars and the possible contribution of globular clusters to the
assembly of the Galactic halo, it is interesting to  point out our
results concerning the comparison between the binary fraction in
clusters and in the field. In particular for P<1000d, we find that the
FG binary fraction is 11.6 \% (the SG is 2.8 \%) while that of the
field population is about 18.8\%.  The FG binary population according
to the scenario outlined above is less affected by dynamical processes
than the SG and its binary fraction is indeed much more similar to
that of the field population. Considering that FG binaries evolved for
one Hubble time in a cluster environment (though less concentrated
than SG binaries) and that the estimated fraction includes long-period
binaries, the difference found between FG and field binary fraction might be 
 due to dynamical effects.  

FG cluster stars share similar chemical properties with halo stars
\citep{gratton12} and multiple-population cluster
formation models  based on self-enrichment predict that the FG
population was initially significantly more massive, was released in
the halo during the cluster early evolution \citep[see e.g.][]{dercole08}, and possibly 
contributed significantly to halo field population. 

Some theoretical and observational efforts to constrain the possible
contribution of clusters to the halo have already been made 
\citep[see e.g.][]{carretta09,martell11,lind15,vesperini10,schaerer11}
however the wealth
of data that will soon be available through several upcoming or
ongoing surveys (APOGEE, Gaia-ESO, GALAH, WEAVE etc) will allow to
address this issue more thoroughly. 

\begin{acknowledgements}
This research was partially supported by the Munich Institute for Astro- and Particle Physics (MIAPP) 
of the DFG cluster of excellence "Origin and Structure of the Universe".
SL, AS, EC, AB acknowledge partial support from PRIN-MIUR 2010-2011 PI Francesca Matteucci
{\it Evoluzione chimica e dinamica della nostra galassia e delle galassie del gruppo locale}
EV acknowledges support by grants NASA-NNX13AF45G and HST-12830.01-A.
\end{acknowledgements}



\end{document}